\documentclass{aa}  

\usepackage{graphicx}
\usepackage{txfonts}
\newcommand{\xmm}{{XMM-{\em Newton} }}
\newcommand{\swift}{{\em Swift }}
\newcommand{\sdsslong}{{SDSS~J120136.02+300305.5 }}
\newcommand{\sdsslongns}{{SDSS~J120136.02+300305.5}}
\newcommand{\sdss}{{SDSS~J1201+30 }}
\newcommand{\sdssns}{{SDSS~J1201+30}}
\newcommand{\swtd}{{SWIFT~J164449.3+573451 }}
\newcommand{\swtdns}{{SWIFT~J164449.3+573451}}
\newcommand{\fluxUnits}{{ergs s$^{-1}$cm$^{-2}$ }}
\newcommand{\fluxUnitsns}{{ergs s$^{-1}$cm$^{-2}$}}
\newcommand{\lumUnits}{{ergs s$^{-1}$ }}
\newcommand{\lumUnitsns}{{ergs s$^{-1}$}}
\newcommand{\brem}{{Bremsstrahlung }}
\begin{document}
   \title{A tidal disruption-like X-ray flare from the 
quiescent galaxy SDSS~J120136.02+300305.5\thanks{Partly based on observations collected at the German-Spanish Astronomical Center, Calar Alto, jointly operated by the Max-Planck-institut f\"ur Astronomie Heidelberg and the Instituto de Astrof\'isica de Andaluc\'ia (CSIC) and observations made with the WHT
operated on the island of La Palma by the Isaac Newton Group in the Spanish Observatorio del Roque de los Muchachos of the Instituto de Astrof\'isica de Canarias.}}

   \subtitle{}

   \author{R.D. Saxton
          \inst{1}
          \and
          A.M. Read\inst{2}
          \and
          P. Esquej\inst{2}\thanks{Now at Centro de Astrobiolog\'ia (CSIC-INTA), E-28850 Torrej\'on de Ardoz, Madrid, Spain}
          \and
          S. Komossa\inst{3,4,5,6}
          \and
          S. Dougherty\inst{7,8} 
          \and
          P. Rodriguez-Pascual\inst{1}
          \and
          D. Barrado\inst{9,10}
          }

   \offprints{R. Saxton}

   \institute{XMM SOC, ESAC, Apartado 78, 28691 Villanueva de la Ca\~{n}ada, Madrid
              , Spain\\
              \email{richard.saxton@sciops.esa.int}
         \and
             Dept. of Physics and Astronomy, University of Leicester, Leicester LE1 7RH, U.K.
         \and
             Technische Universit\"at M\"unchen, Fakult\"at f\"ur Physik, James-Franck-Strasse 1/I, 85748 Garching, Germany
         \and
             Excellence Cluster Universe, TUM, Boltzmannstrasse 2, 85748 Garching, Germany
         \and
             Max Planck Institut f\"ur Plasmaphysik, Boltzmannstrasse 2, 85748 Garching, Germany
         \and
             National Astronomical Observatories, Chinese Academy of Sciences, 20A Datun Road, Beijing, 100012, China
         \and
            National Research Council, Herzberg Institute of Astrophysics, DRAO, Box 248, Penticton, BC, V2A 6J9
         \and
             Dept of Physics and Astronomy, University of Calgary, 2500 University Dr. NW., Calgary, AB T2N 1N4 
         \and
             Centro de Astrobiolog\'ia Depto. Astrofisica (INTA-CSIC), ESAC campus, Apartado 78, 28691 Villanueva de la Ca\~{n}ada, Spain
         \and
             Calar Alto Observatory, Centro Astron\'omico Hispano Alem\'an, Almer\'ia, Spain
        }


 
  \abstract
   {}
   {The study of tidal disruption flares from galactic nuclei 
has historically been hampered by a
lack of high quality spectral observations taken around the peak 
of the outburst.
Here we introduce the first results from a program designed to identify
tidal disruption events at their peak by making near-real-time comparisons 
of the flux seen in \xmm slew sources with that seen in ROSAT.}
   {Flaring extragalactic sources, which do not appear to be AGN, are monitored with \swift and \xmm to track their temporal and spectral evolution.
Timely optical observations are made to monitor the reaction of circumnuclear
material to the X-ray flare.}
   {\sdsslong was detected in an \xmm slew from June 2010 
with a flux 56 times higher than an upper limit from ROSAT, corresponding
to L$_{X}\sim3\times10^{44}$ \lumUnitsns. It has the optical spectrum of a
quiescent galaxy (z=0.146). Overall the X-ray flux has evolved 
consistently with the canonical t$^{-5/3}$ model, expected for 
returning stellar debris, fading by a factor $\sim 300$ over 300
days. In detail the source is very 
variable and became invisible to \swift between 27 and 48 days after 
discovery, perhaps due to self-absorption. 
The X-ray spectrum is soft but is not the expected tail of 
optically thick thermal emission.
It may be fit with a \brem or double-power-law model and is seen to
soften with time and declining flux. Optical spectra taken 12 days and 11 
months after discovery indicate a deficit of material in the broad line
and coronal line regions of this galaxy, while a deep radio non-detection 
implies that a jet was not launched during this event. 
}
   {}


   \maketitle
%

\section{Introduction}

It has been suggested that the flare of radiation produced by
the tidal disruption of a star is a
unique signature of the presence of otherwise dormant
supermassive black holes (SMBH) in the nuclei of galaxies (e.g., \cite{Rees88}).
The accretion of tidally disrupted stellar debris
may significantly contribute to the low-luminosity end of the
AGN luminosity function (\cite{Milosavljevic}).
To date, several candidate tidal
disruption events (TDE)
have been identified, based on a luminous flare seen at soft X-ray
(e.g., \cite{Komossa99};
 \cite{Greiner}; \cite{Esquej07}; \cite{Cappelluti09}), 
UV (e.g., \cite{Gezari06};
\cite{Gezari08})
or optical (e.g., \cite{Komossa08}; \cite{Cenko}; \cite{vanVelzen11})
wavelengths.
The first X-ray tidal disruption flares were identified in the course
of the ROSAT all-sky survey (\cite{Komossa99}; \cite{Komossa99b}; 
\cite{Grupe99}; \cite{Greiner}).
These events all shared very similar properties,
including high peak luminosities and very soft spectra characterized by thermal
emission at 40-100 eV, and, in the case of NGC~5905, 
a decline law scaling approximately as
$t^{-5/3}$ (see \cite{Komossa02} for a review). 
Subsequent \xmm and
{\em Chandra} observations of these sources, taken 10 years after their
discovery showed fading by a factor 200-6000 and a
remnant spectrum which had hardened (\cite{Halpern04}; \cite{Komossa04};
 \cite{Vaughan}).

The rate of disruption of stellar objects by the central 
SMBH of a galaxy also has the potential to give much information about stellar
dynamics in galactic nuclei. 
Donley et al. (2002) estimated an X-ray tidal disruption rate     
 of $\sim10^{-5}$yr$^{-1}$galaxy$^{-1}$ in rough agreement with
theoretical predictions (e.g. \cite{Magorrian}; \cite{WangMerritt}).

Basic TDE theory predicts that debris from the star will fall back
to the SMBH, shock against itself and form
an optically thick torus of material close to the innermost stable orbit
(\cite{Rees88}; \cite{Cannizzo}). This material will be accreted and
replenished
at a rate of $t^{-5/3}$ (\cite{Phinney}).
More recent simulations imply that the initial decrease in flux may be
flatter (\cite{Lodato09})
and that the flare could remain close to its peak for many months before
tending to a $t^{-5/3}$ decline.  

Optical reprocessing of the X-ray flare from surrounding
matter will to first order be similar whether the flare was generated
from a TDE or an active galactic nucleus (AGN). Any existing broad line
gas will be illuminated within a
few days or tens of days.
Changes in broad line 
emission have previously been seen in the active galaxy IC3599 (\cite{Brandt},
\cite{Grupe95a})
and in the galaxies SDSS~J095209.56+214313.3 (\cite{Komossa08}) and 
SDSS~J074820.67+471214.3 (Wang et al. 2011).  
The coronal line region (CLR) will also be illuminated,
dramatically in the case of SDSS~J0952+2143, a few months
to a few years after the flare.

To investigate the behaviour of the flare around the peak it is necessary
to catch one in the act. To this end we have initiated a program to
monitor sources found in \xmm slews (\cite{Saxton08})
and compare their flux quickly
with that seen in ROSAT. Extragalactic sources, which show a strong increase
in flux, and are not known AGN, then become candidates for a follow-up
program.
Such sources are very rare. We report here on the first fruit from this
program; the galaxy \sdsslongns, observed
during an \xmm slew with a flux 56 times greater than the upper limit
from the ROSAT all-sky survey (RASS; \cite{Voges}).

The paper is structured as follows:
in Section 2 we discuss the flare detection and source identification.
In Section 3 we review follow-up observations of the optical spectrum,
in Section 4 we present follow-up X-ray and UV observations
and describe the data analysis and in Section 5 we detail follow-up 
radio observations. Section 6 assesses whether the flare
characteristics can be explained by an AGN or a TDE and discusses the
source properties within the TDE model. The paper is summarised in
Section 7.

A $\lambda$CDM cosmology with ($\Omega_{M},\Omega_{\Lambda}$) = (0.27,0.73)
and  $H_{0}$=70 km$^{-1}$s$^{-1}$ Mpc$^{-1}$ has been assumed throughout.

\section{X-ray flare identification}
During the slew 9192300005, performed on June 10$^{th}$ 2010, \xmm detected
a source, XMMSL1~J120135.7+300306, with an EPIC-pn, medium filter,
 0.2--2 keV count rate of 
$2.1\pm0.5$ count s$^{-1}$. For a typical 
theoretical tidal disruption spectrum of a black body of temperature 70 eV
and the Galactic column for this sky position of $1.4\times10^{20}$ cm$^{-2}$
(\cite{Kaberla}),
 this corresponds to an unabsorbed flux of 
F$_{0.2-2.0}=3.1\times10^{-12}$ \fluxUnitsns.
We calculate a 2-sigma upper limit from the RASS at this position of
0.03 count s$^{-1}$ (see \cite{Esquej07} for a description of the upper limit
calculation), corresponding to F$_{0.2-2.0}\leq5.4\times10^{-14}$ \fluxUnitsns, 
using the same spectral model; a factor 56 below the \xmm 
slew value.

The source position was also observed during two earlier slews 
yielding weak 2-$\sigma$
upper limits of $<0.67$ count s$^{-1}$ in 2003-11-24 and $<1.2$ count s$^{-1}$ in 2007-06-11. 

\begin{figure}
\centering
\rotatebox{0}{\includegraphics[height=7cm]{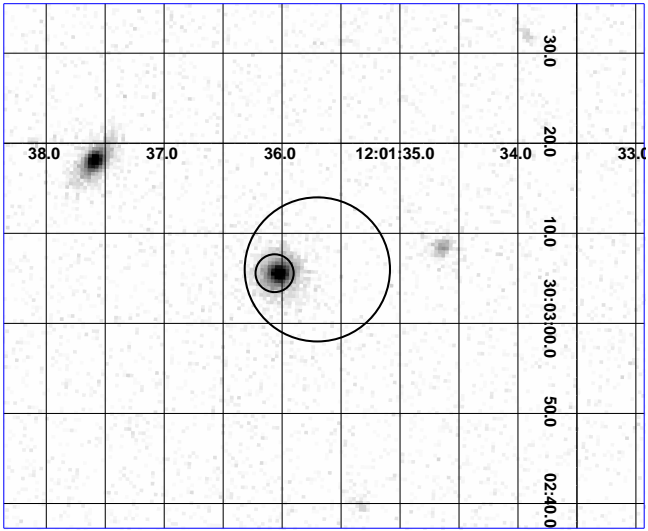}}
\caption[SDSS image]
{ \label{fig:sdssimage} An r-filter image of the galaxy from the SDSS survey, 
(limiting magnitude r$\sim23$), taken on 2004-12-13, 
 shown with the \xmm slew error circle (8 arcsecond radius) and UVOT-enhanced
\swift error circle (2.1 arcsecond radius; see text) centred on the detections.
}

\end{figure}

The error on slew source positions is dominated by the systematic error in 
reconstituting the satellite attitude and has been measured as 8 arcseconds
(1 sigma) for an ensemble of sources (\cite{Saxton08}).
Within this error circle there is only one catalogued bright source 
(Fig.~\ref{fig:sdssimage}),
\sdsslong (hereafter \sdss), with r=17.93, g=18.95, u=20.78, i=17.50, z=17.18;
a galaxy located at 4 arcseconds from the \xmm position, with a photometric
redshift z=0.128 (SDSS-DR7; \cite{sdss7}). This source is coincident with 
2MASS 12013602+3003052: J=$16.43\pm0.11$, H=$16.13\pm0.18$, 
K=$15.43\pm0.19$.
The galaxy itself appears to be extended, round, face-on and with 
little structure.

A crude analysis may be performed on the 19 photons in the 
slew spectrum to investigate the gross spectral properties of the 
detection. Detector matrices are calculated, taking into account 
the transit of the
source across the detector, using a technique
outlined in Read et al. (2008). The source is soft, with a power-law 
slope $\sim 3$
or black-body temperature $\sim 0.1$ keV, assuming no intrinsic
absorption above the Galactic value.

\section{Optical observations}
\begin{figure}
\centering
\rotatebox{-90}{\includegraphics[height=9cm]{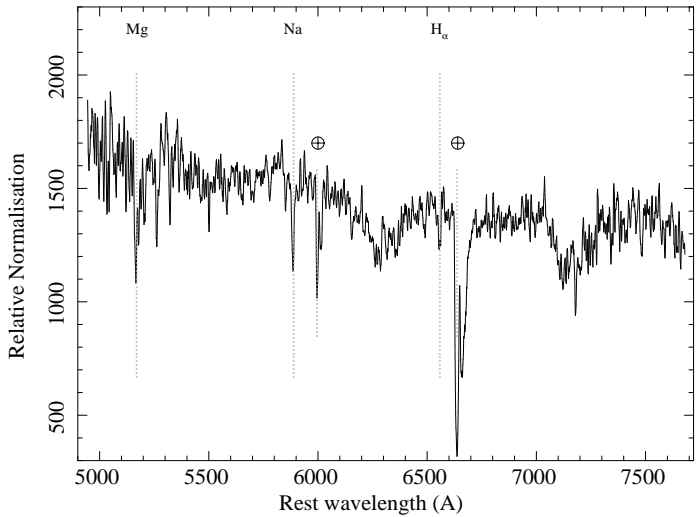}}
\caption[Calar Alto spectrum]
{ \label{fig:calaltspec} Calar Alto 3.5m / TWIN spectrum of \sdss 
taken on 2010-06-22, 12 days after discovery, 
smoothed by a factor 4. $\oplus$ denotes the position
of the Tellurium atmospheric absorption lines.}

\end{figure}

Director's time was granted on the 3.5 m / TWIN spectrograph at Calar Alto
to observe the source on 2010-06-22, 12 days after the discovery. 
A 1 hour exposure in the 
red band (5666 - 8801 \AA\ )
was made, under photometric conditions with 1.5$\arcsec$ seeing.
The resultant spectrum, with a dispersion of 1.6 \AA\ per pixel  
and signal-to-noise ratio of 15, contains no emission lines
(Fig.~\ref{fig:calaltspec}).
The absorption lines Mg~I $\lambda$5167,5173,5184 and 
Na~I $\lambda\lambda$5890,5895 can be identified
and give a redshift of $0.146\pm{0.001}$, a little higher than the
photometric redshift obtained from the SDSS filters.

\begin{figure}
\centering
\rotatebox{-90}{\includegraphics[height=9cm]{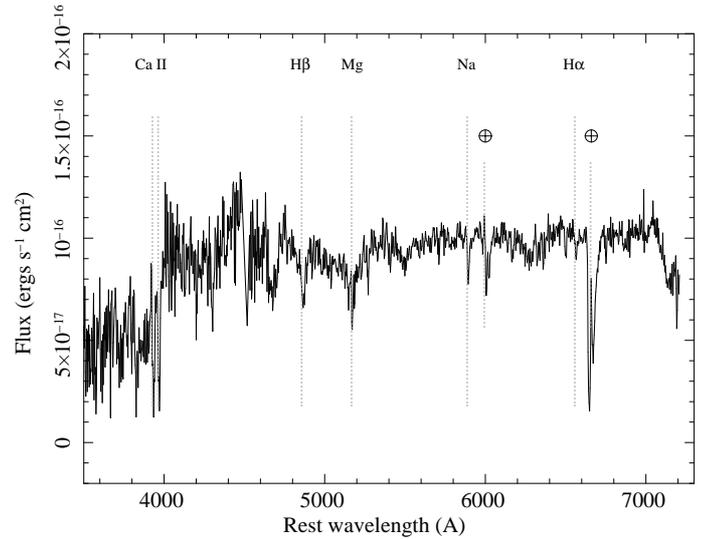}}
\caption[WHT spectrum]
{ \label{fig:whtspec} WHT/ISIS spectrum of \sdss taken with the 
R316R, R300B grisms on 2011-05-11, 11 months after discovery, 
smoothed by a factor 4. 
$\oplus$ denotes the position
of the Tellurium atmospheric absorption lines.} 

\end{figure}

A further service-time spectrum was taken on the WHT, using the ISIS 
spectrograph with the R316R and R300B grisms (Fig.~\ref{fig:whtspec}). 
This second spectrum, taken on
2011-05-11, 11 months after the flare, was designed to extend the 
coverage to blue wavelengths and catch the possible
 light echo 
of the flare from circumnuclear material. A spectacular example of this effect
was seen in SDSS~J0952+2143 (\cite{Komossa08}) where coronal lines 
were illuminated $\sim$one year after a strong nuclear flare. 
The exposure time was 30 minutes giving a signal-to-noise ratio of $\sim5$
in the blue band and 15 in the red band with a dispersion of 
$\sim0.8$ \AA\ per pixel.
Once again no lines were detected in emission. H${\alpha}$ and 
H${\beta}$ are seen in absorption and together with Mg~I 
$\lambda$5167,5173,5184, Na~I $\lambda\lambda$5890,5895, Ca~II H and
Ca~II K absorption lines give a redshift of $0.144\pm{0.002}$. 
After modelling and subtraction of an old (11Gyr) stellar population template
(\cite{Bruzual}) no optical emission lines are seen.
The [OIII] $\lambda$5007 emission may be constrained to 
F$_{[OIII]}\leq5\times10^{-17}$ \fluxUnitsns; $L_{[OIII]}\leq4\times10^{39}$ 
\lumUnitsns.
From bolometric correction factors of Lamastra et al. (2009) we infer an 
upper limit to the total luminosity of any persistent emission
of  $L_{bol}\leq3\times10^{41}$ ergs s$^{-1}$.
 
\section{X-ray follow-up observations}
An X-ray monitoring program was initiated with \swift to 
follow the evolution of the source flux and spectrum. Snapshot 3ks observations
were made once a week with the \swift-XRT (\cite{Burrows05}) in photon
counting mode and the UV optical telescope (UVOT; \cite{Roming}),
using the filter of the day, until the source left the visibility window in 
August 2010. The \swift-XRT observations have been analysed following the 
procedure outlined in Evans et al. (2009) and the UVOT data have been 
reduced as described in Poole et al. (2008). 
An accurate position for the source in the \swift-XRT field can be determined
by matching the UVOT field of view with the USNO-B1 catalogue and registering
the XRT field accordingly (\cite{Goad}). The source is coincident with
the galactic nucleus (see Fig.~\ref{fig:sdssimage}).
In parallel, a 30ks \xmm pointed observation was triggered
on 2010-06-22 (coeval with the Calar-Alto/Twin optical spectrum) 
and again on 2010-11-23.
The 2010-11-23 observation was strongly affected by radiation and was repeated
on 2010-12-23. In each observation the EPIC-pn, MOS-1 and MOS-2 cameras 
were operated in full frame mode with the thin1 filter in place. 
The source was too
faint for statistically significant data to be collected from the
reflection grating spectrometers. 
The XMM data were analysed with the \xmm Science Analysis System 
(SAS v11.0.0; \cite{Gabriel}). Light curves were extracted from the 
observations and searched for periods of high background flaring.
These periods were excluded from further data analysis and the total
amount of remaining exposure time is listed in Table.~\ref{tab:xobs}.
Source events were extracted from a circle of radius 25(20) arcseconds 
for the EPIC-pn, MOS-2 (MOS-1) cameras, using patterns 0-4 (EPIC-pn) 
and patterns 0-12 (MOS-1, MOS-2). Background events were extracted from 
a source-free region on the same CCD.
A list of all X-ray observations is given 
in Table~\ref{tab:xobs}.

\begin{center}
\begin{table*}[ht]
{\small
\caption{X-ray observation log of \sdss}
\label{tab:xobs}      
\hfill{}
\begin{tabular}{l c l l l}
\hline\hline                 
Mission$^{a}$ & Date & Exp time$^{b}$ &  Count rate$^{c}$ & Flux$^{d}$ \\
              &      &   (s)    &    (count s$^{-1}$) & ($10^{-12}$\fluxUnitsns) \\
\\
ROSAT  & 1990  & 500  & $<0.0068$ &  $<0.054$  \\
XMM slew  &   2010-06-10  & 8.9 &  $2.1\pm{0.5}$ & $4.7^{+3.4}_{-2.0}$   \\
SWIFT  & 2010-06-20       &  2996 & $0.014\pm{0.002}$ & $0.56\pm{0.26}$ \\
XMM pointed & 2010-06-22  &  18865 &  $0.37\pm{0.004}$ & $0.81\pm{0.07}$ \\
SWIFT  & 2010-06-30  & 3252 &   $0.041\pm{0.004}$ & $1.4\pm{0.4}$ \\
SWIFT  & 2010-07-07  & 3580 & $<0.00084$ & $<0.030$ \\
SWIFT  & 2010-07-14  & 2780 &  $<0.0011$ & $<0.039$ \\
SWIFT  & 2010-07-21  & 3065 &  $<0.0010$ & $<0.035$ \\
SWIFT  & 2010-07-28  & 505  &  $<0.0059$ & $<0.21$ \\
SWIFT  & 2010-10-24  & 3189 &  $0.0037\pm{0.0013}$ & $0.13\pm{0.05}$ \\
XMM pointed &   2010-11-23      &  10697 & $0.079\pm{0.003}$ & $0.19\pm{0.03}$\\
XMM pointed &   2010-12-23      &  11303 & $0.013\pm{0.001}$ & $0.04\pm{0.02}$\\
SWIFT  & 2011-03-25    &  2285    & $<0.0013$  & $<0.046$\\
SWIFT  & 2011-04-01    & 2912    &   $<0.0010$ & $<0.035$\\
SWIFT  & 2011-04-11    & 2688    &  $<0.0011$  & $<0.039$ \\
\hline                        
\end{tabular}}
\hfill{}
\\
\\
$^{a}$ \xmm, EPIC-pn camera: slew observation performed in {\em full frame}
mode with
the Medium filter; pointed observations performed in {\em full frame} mode with
the thin1 filter. \swift-XRT in pc mode. \\
$^{b}$ Useful exposure time after removing times of high background flares. \\
$^{c}$ Count rate in the band 0.2--2 keV. \\
$^{d}$ Unabsorbed flux, $F_{0.2-2~keV}$ calculated with the best fit 
\brem model (or kT=300 eV for the upper limits; see Section 4.3) and 
Galactic absorption of $1.4\times10^{20}$cm$^{-2}$ \\
\end{table*}
\end{center}

%
%

\subsection{X-ray light curve}
\begin{figure}
\centering
\rotatebox{-90}{\includegraphics[height=9cm]{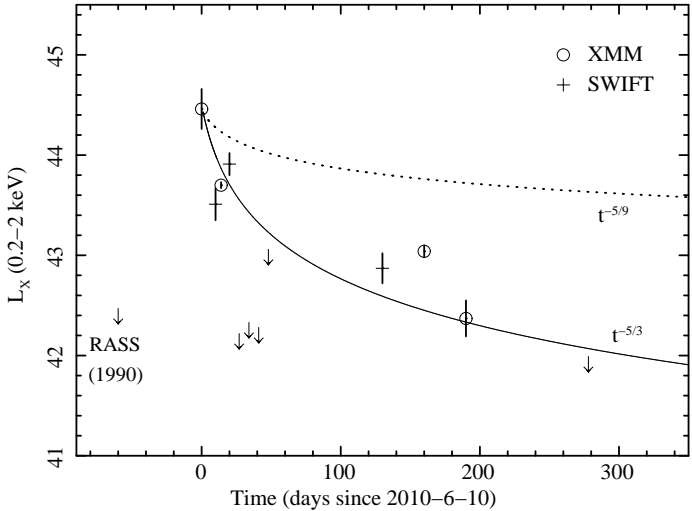}}
\caption[\sdss light curve]
{ \label{fig:lcurve} The 0.2--2 keV X-ray light curve of \sdss. \xmm slew
and pointed observations: circle, \swift-XRT: cross. All upper limits are
from \swift except the RASS point. The latest upper limit point is calculated
by combining the last three \swift observations from 2011-03-25, 2011-04-01
and 2011-04-11. The lines represent t$^{-5/3}$ (solid) and 
t$^{-5/9}$ (dotted) declines.}

\end{figure}

In figure~\ref{fig:lcurve} we show the X-ray light curve 
for SDSS~J1201+30. 
The source is highly variable, with a factor 50 drop in flux between 
2010-06-30 and 2010-07-07. We have superimposed two fiducial decay curves
onto the X-ray light curve: a canonical t$^{-5/3}$ evolution related to
the rate of return of tidal debris and a t$^{-5/9}$
decline which is that predicted for EUV emission from an expanding wind 
(\cite{Strubbe09}). While the strong variations make it impossible
to fit a smooth curve through the data points, the overall decay
appears to be steeper than the t$^{-5/9}$ law but not inconsistent
with a t$^{-5/3}$ decline. 
If the latter is correct, then the source will be 
emitting at L$_{X}\sim10^{41-42}$ \lumUnits for the next 
1-2 years and will continue 
to be visible to \xmm or {\em Chandra}. A shallower initial 
light curve is predicted by detailed modelling of the stellar internal 
density (\cite{Lodato09}).
With the strong observed variability it is impossible to say what the
shape of the underlying early-phase intrinsic light curve was in this case.
Qualitatively the light curve
bears a striking resemblance to that of SWIFT~J164449.3+573451 
(\cite{Burrows11}), which shows variations by up to two decades in flux 
over a few days, superimposed on a gradual downward trend. In that
case the emission is believed to be caused
by a relativistic jet which has been instantiated by the sudden accretion of
material from a tidally disrupted star.

\subsubsection{Short-term X-ray variability}

\begin{figure}
\centering
\rotatebox{-90}{\includegraphics[height=9cm]{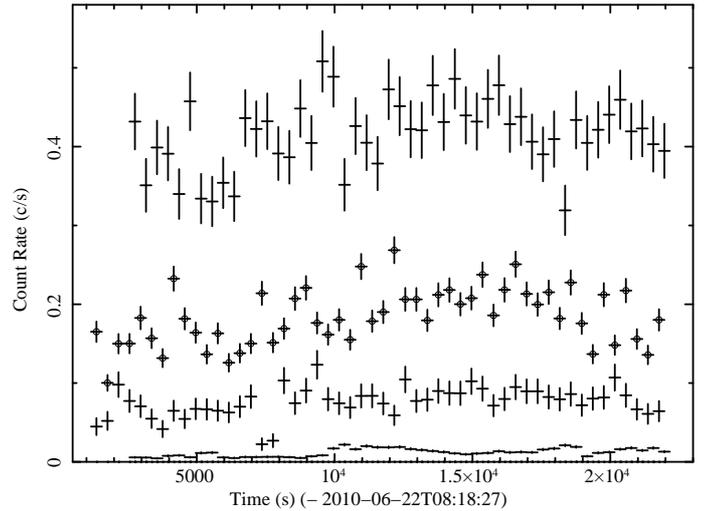}}
\caption[\sdss short-term light curve]
{ \label{fig:lcurve_short} The background-subtracted, exposure corrected, 
EPIC-pn (top), MOS-1 (middle) and MOS-2
(next to bottom) 0.16-1.0 keV light curves from the first 
\xmm pointed observation of 2010-06-22, together with the exposure 
corrected, EPIC-pn background (bottom),
which reaches a maximum of 0.022 count s$^{-1}$. The MOS-1 curve is scaled by a factor 2.5 for ease of viewing.}

\end{figure}

The first \xmm pointed observation was analysed to 
search for rapid variability.
A 0.16-1 keV light curve was extracted 
and binned into 400 second time bins. 
The light curve was exposure corrected and background subtracted using
the XMM-SAS task {\em epiclccorr}.
A K-S test on the EPIC-pn light curve (Fig.~\ref{fig:lcurve_short})
gives a probability of the source being variable of 98.7\%.
The shortest variability scale is $\sim4000$ s 
with an amplitude of $\sim50$\%, which can be seen from 6--10 ks 
in Fig.~\ref{fig:lcurve_short}.
The second \xmm pointed observation shows similar variability
albeit with less statistical significance.

\subsection{UV light curve}

During the three pointed \xmm observations, the optical monitor (OM)
cycled between the {\it U}, {\it UVW1} and {\it UVM2} filters. 
\swift-UVOT observations were
performed with the filter of the day; once with the {\it uv} filter and
otherwise with the {\it uvw1, uvw2} and {\it uvm2} filters in place.
The galaxy is faint at UV wavelengths
 and was only detected in the XMM-{\it U}, \swift-{\it uv} 
and XMM-{\it UVW1} filters. 
The XMM-{\it U} (3480\AA) and XMM-{\it UVW1} (2950\AA) fluxes are 
constant between the observations (Tab.~\ref{tab:uvflux}). We can place
a 2-$\sigma$ upper limit on any decrease in the XMM-U count rate
between the 2010-06-22 and 2010-12-23 \xmm observations
of 0.025 count s$^{-1}$. The low UV flux argues against the galaxy 
containing a significant young stellar population.

\begin{table}
\caption{\xmm optical monitor UV measurements of \sdss}
\label{tab:uvflux}      
\begin{center}
\begin{tabular}{l c c c}
\hline\hline                 
Date & \multicolumn{3}{c}{Flux$^{a}$} \\
      &    U       &  UVW1 & UVM2 \\
\\
2010-06-22  & $2.5\pm{0.4}$ & $1.0\pm{0.4}$ & $<1.7$ \\
2010-11-23  & $2.5\pm{0.3}$ & $0.7\pm{0.3}$ & $<1.0$ \\
2010-12-23  & $2.6\pm{0.3}$ & $0.7\pm{0.3}$ & $<1.1$ \\
\hline                        
\end{tabular}
\\
\end{center}
$^{a}$ Observed flux, units of 10$^{-17}$ erg cm$^-2$ s$^{-1}$ \AA$^{-1}$
 for the 
\xmm OM {\it U} (3480\AA), {\it UVW1} (2950\AA) and
{\it UVM2} (2340\AA) filters. Errors are 1-$\sigma$.

\end{table}

\subsection{X-ray spectrum}
\begin{figure}
\centering
\rotatebox{-90}{\includegraphics[height=9cm]{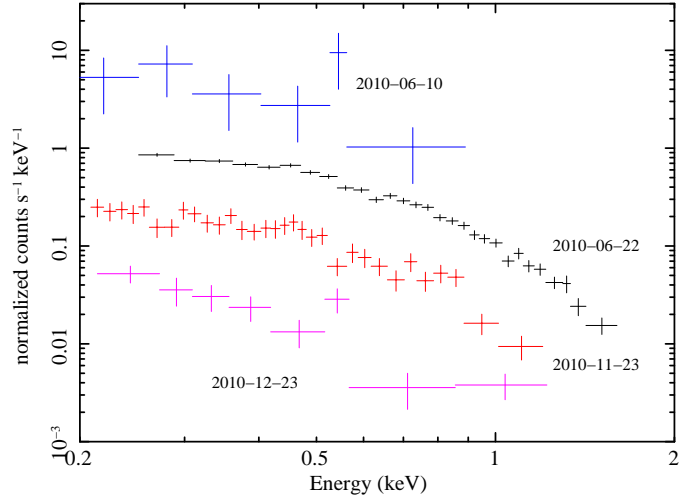}}
\caption[XMM spectra]
{ \label{fig:cntspec} \xmm, EPIC-pn spectra of \sdss from the slew
survey (medium filter) and the three pointed observations (thin1 filter).}
\end{figure}

Despite the factor 100 drop in count rate between the \xmm slew 
observation and the \xmm pointed observation of 2010-12-23, 
taken 190 days later, there is little 
change in the shape of the spectrum (Fig.~\ref{fig:cntspec}). 

Spectra were produced for the three EPIC cameras from the 2010-06-22 \xmm
observation and fit with XSPEC
(version 12.6.0).  Fits were performed simultaneously on the ungrouped
spectra, using the Cash statistic (\cite{Cash}), over the energy range  
0.2 - 2keV, using a constant to account for
the small differences in normalisation between the instruments.

\begin{figure}
\centering
\rotatebox{-90}{\includegraphics[height=9cm]{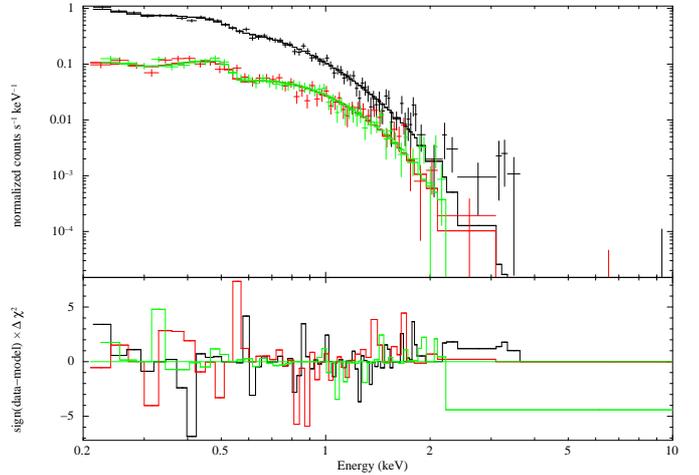}}
\caption[Spectral fit to the spectra from the 2010-06-22 \xmm observation]
{ \label{fig:obs1spec} A fit to the \xmm EPIC-pn (black), MOS-1 (red / grey)
and MOS-2 (green / light grey) spectra of \sdss from the 
2010-06-22 observation with a rest-frame model of a \brem of kT=390 eV and 
an edge at 660 eV.}
\end{figure}

The overall spectrum is relatively broad and can not be modelled with 
a black body or disk black body model (diskbb in {\it xspec}; 
Tab.~\ref{tab:specfits}). 
A power-law does not fit the data well, either using extra intrinsic
neutral or ionized absorption or an additional soft component.
The spectral shape is modelled reasonably well by either 
a \brem of temperature kT=390 eV (C/dof=1.07) or a broken power-law
model (C/dof=1.04) if an edge is included at an energy of
$\sim 0.66$ keV in the source rest frame (Fig.~\ref{fig:obs1spec}). 
The observed edge energy is roughly 
that of neutral Oxygen at zero redshift. It is possible that there could be
an excess of Oxygen in the line of sight in our galaxy but it would
need to have an abundance 5 times higher than solar. 
We note that the spectra show no evidence for line emission. If an APEC
model, with kT=390 eV, is fitted then the elemental abundances are 
limited to 0.5\% of the solar values.

The \xmm pointed observations of 2010-11-23 and 2010-12-23
were analysed in the
same way. EPIC spectra produced from the 2010-11-23 
observation are fit well with a \brem of temperature kT=290 eV.
There is no requirement for extra absorption, although formally the
edge seen in the first observation at 655 eV
with $\tau=0.3$ is not excluded. The 2010-12-23 observation (EPIC-pn
only) may be fit with
a \brem of kT=180 eV and again an edge does not improve the fit. 
A broken power-law 
model is also acceptable for these two observations.

The \swift-XRT observations of 2010-06-20 and 2010-06-30 have low
statistics but may be used to constrain the \brem temperature or
the $\Gamma$ for a power-law fit. The observation of 2010-06-30 has
the hardest spectrum of all.

The best fit flux from the \xmm slew observation is 
$4.7^{+3.4}_{-2.0}\times10^{-12}$ \fluxUnits corresponding to 
$L_{X}=2.8^{+2.0}_{-1.2}\times10^{44}$ \lumUnits.
Extrapolating over the EUV band gives a bolometric luminosity 
of $L_{bol}=5-14\times10^{44}$ \lumUnitsns.

\begin{figure}
\centering
\rotatebox{-90}{\includegraphics[height=9cm]{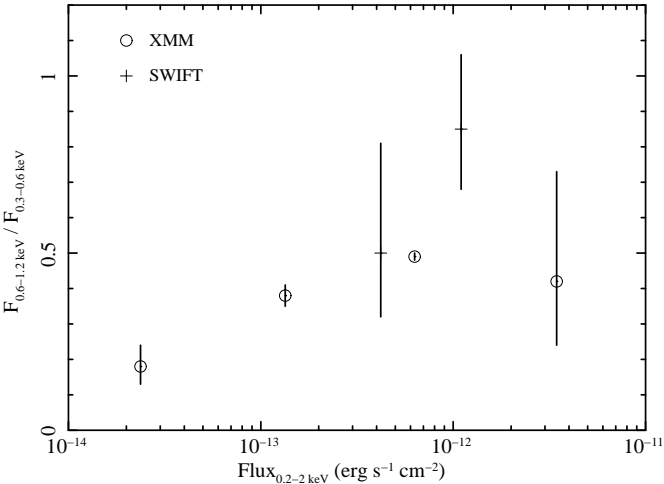}}
\caption[Hardness ratios]
{ \label{fig:hratio} The 0.6--1.2 keV to 0.3--0.6 keV flux ratio as
a function of the (absorbed) soft X-ray (0.2--2 keV) flux for
observations with sufficient statistics to constrain the flux
ratio.}
\end{figure}

The spectrum is seen to get softer as the flux declines (Fig.~\ref{fig:hratio}).
The same evolution was seen in the spectrum of \swtd which also 
softens with decreasing flux (\cite{Burrows11}). 
This differs from previous TDE flares which showed a spectral hardening with 
time in observations of candidates taken several years after the peak 
of the event (\cite{Halpern04}; \cite{Komossa04}; \cite{Vaughan}). 
This may be telling us that the later emission (after $\sim2$ years)
is caused by a different
process, for example accretion from a slim disk. A summary of fit parameters
is presented in Tab.~\ref{tab:specfits}.

\begin{table*}[ht]
{\small
\caption{Spectral fits to \xmm and \swift observations of \sdss}
\label{tab:specfits}      
\begin{center}
\begin{small}
\begin{tabular}{c c c c c c c | c l l |r}
\hline\hline                 
 \multicolumn{7}{c}{emission model} & \multicolumn{3}{c}{intrinsic absorption} & C/dof  \\
\\
Plaw  &  \multicolumn{3}{c}{Bkn-power} & zbrem  &  \multicolumn{2}{c}{Diskbb}  &  ztbabs  &  \multicolumn{2}{c}{Edge}  & \\
$\Gamma$   & $\Gamma_{1}$ & $\Gamma_{2}$ & $E$ (keV) &  kT (keV)$^{a}$ & Norm & $E$ (keV)$^{a}$ & $N_{H}$ & E (keV)$^{a}$ & $\tau$ & \\
\hline
\multicolumn{11}{c}{\xmm observation 1 - 2010-06-22} \\
\hline
-   & -         &  -      &  -   &    - &  22000 &  0.07$^{*}$ & - & - & - & 11396/571  \\
$3.38\pm{0.04}$   & -         &  -      &  -   &    - &  -  &  - & - & - & - & 775/570  \\
3.97   & -         &  -      &  -   &    - &  -  &  - & $5.1\pm{1.0}\times10^{20}$   &  -  &  - & 648/569  \\
3.3   & -         &  -      &  -   &    - &  0  &  0.07$^{*}$ & - & - & - & 775/568  \\
-   & 2.6 &  4.5      &  0.7   & -  &   -  &  - & - & - & - & 620/568 \\
-   & 2.6 &  4.5      &  0.7   & -  &   -  &  - & - & $0.67\pm{0.03}$ & $0.4\pm{0.07}$ & 589/566 \\
-   & - &  - &  -   &  $0.38\pm{0.01}$  &  -  & - & - & - & - & 634/570 \\
-   & - &  - &  -   &  $0.39\pm{0.01}$  &  -  & - & - & $0.655\pm{0.035}$ & $0.3\pm{0.1}$ & 606/568 \\
\hline
\multicolumn{11}{c}{\xmm observation 2 - 2010-11-23} \\
\hline
$3.69\pm{0.12}$   & -         &  -      &  -   &    - &  -  &  - & - & - & - & 586/595  \\
-   & 2.6 &  6.0      &  0.7   & -  &   -  &  - & - & - & - & 524/593 \\
-   & 2.6 &  5.7      &  0.65   & -  &   -  &  - & - & 0.655$^{*}$ & 0.3$^{*}$ & 520/591 \\
-   & - &  - &  -   &  $0.29\pm{0.02}$  &  -  & - & - & - & - & 512/595 \\
-   & - &  - &  -   &  $0.29\pm{0.02}$  &  -  & - & - & 0.655$^{*}$ & 0.3$^{*}$ & 516/595 \\
\hline
\multicolumn{11}{c}{\xmm observation 3 - 2010-12-23} \\
\hline
$4.24\pm{0.45}$   & -         &  -      &  -   &    - &  -  &  - & - & - & - & 212/359  \\
-   & 3.1 &  7.0      &  0.54   & -  &   -  &  - & - & - & - & 205/357 \\
-   & - &  - &  -   &  $0.18^{+0.06}_{-0.03}$  &  -  & - & - & - & - & 205/359 \\
\hline                        
\multicolumn{11}{c}{\swift observation 1 - 2010-06-20} \\
\hline
$3.69\pm{0.75}$   & -         &  -      &  -   &    - &  -  &  - & - & - & - 
& 12/70  \\
-   & - &  - &  -   &  $0.36^{+0.18}_{-0.09}$  &  -  & - & - & - & - & 11/70 \\
\hline
\multicolumn{11}{c}{\swift observation 2 - 2010-06-30} \\
\hline
$2.80\pm{0.38}$   & -         &  -      &  -   &    - &  -  &  - & - & - & - & 52/62  \\
-   & - &  - &  -   &  $0.63^{+0.23}_{-0.14}$  &  -  & - & - & - & - & 52/62 \\
\hline
\end{tabular}
\\
\end{small}
\end{center}
All fits include absorption by the Galactic column ($N_{H}=1.4\times10^{20}$cm$^{-2}$). Errors are 90\% confidence. \\
$^{a}$ Temperature or energy in the source rest frame. \\
$^{*}$ Frozen parameter \\
}
\end{table*}

\section{Radio observations}
A search for radio emission associated with SDSSJ1201+30 was made on
2011-09-15 with the NRAO Very Large Array (VLA) in New Mexico, USA
using Director's Discretionary Time. Observations were completed at
1.388 (L band), 4.832 (C band) and 8.332 GHz (X band), using 256 MHz
of bandwidth. Observations of the target were made between observations
of the nearby phase-reference calibrator J1207+2754.  The flux scale
was established by an observation of J1331+305 (3C286).

Editing, calibration, imaging, deconvolution and analysis were
completed using the NRAO AIPS software package.  The fluxes
established for J1207+2754 were 0.452, 0.357, and 0.352 Jy at 1.4, 4.8
and 8.3 GHz respectively. Images at each frequency were made of an
area extending at least 30 arcseconds around the coordinates of
\sdssns, and weighted in order to increase the detection
sensitivity of the imaging. No emission was detected at the location
of \sdssns, leading to $3\sigma$ upper limits on the radio flux
of $201, 135$ and $108$ $\mu$Jy at 1.4, 4.8 and 8.3 GHz respectively.

\begin{figure}
\centering
\rotatebox{90}{\includegraphics[height=10cm]{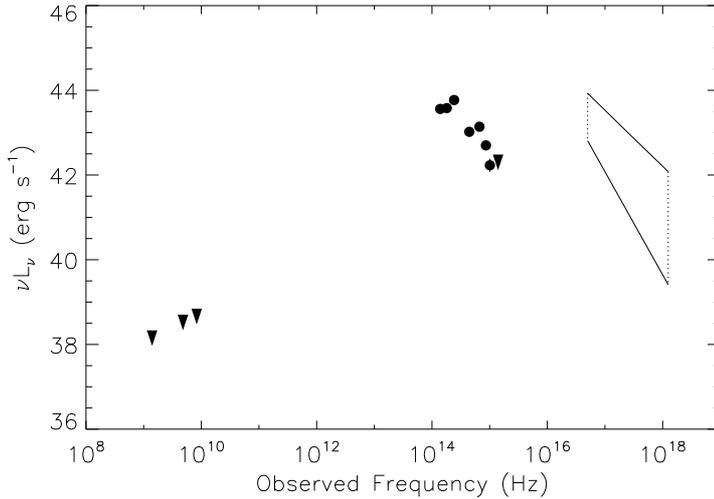}}
\caption[Spectral energy distribution]
{\label{fig:sed} The spectral energy distribution of \sdssns, consisting
of VLA 3-sigma upper limits, extinction corrected 2MASS and WHT data,
UV data from the \xmm OM (2010-06-22) and a power-law fit to the
\xmm X-ray data of 2010-06-22 (upper solid line) and 
2010-12-23 (lower solid line).
} 

\end{figure}

The radio measurements of \sdss are plotted in Fig.~\ref{fig:sed}
together with multi-wavelength observations.

\section{Discussion}

\subsection{High-amplitude variability of AGN}

Before we model in detail the spectral and temporal evolution of this
extragalactic flare we need to assess whether it has been caused by a
stellar disruption, an AGN event or a combination of the two.

X-ray variability by a factor $>$100 is very rare in AGN 
but has been detected in a  few cases, 
so the {\em amplitude} of variability alone can not distinguish between 
the two scenarios in the case of \sdss.

Below, we briefly summarize the characteristics of high variability AGN,
and of some previously detected TDEs.
In the next Section, we then compare them with \sdss. 

Variability of AGN by factors of a few to 20 is not uncommon,
and is typically traced back to changes in the line-of-sight 
(cold or ionized) absorption (e.g., \cite{Risaliti05}), 
or can be caused by emission and reflection mechanisms 
(e.g., \cite{Fabian11}). In rare cases, this variability
has been semi-periodic (\cite{Otani};  \cite{Boller07}).  
Semi-periodicity is also predicted from 
periodic interaction of a compact object with the
accretion disk, which can cause recurrent flares in AGN. Large thermal \brem
flares seen at optical wavelengths in OJ~287 may be caused by
a secondary massive black hole passing through the accretion disk
of the primary (\cite{LehtoValt}).

Variability exceeding a factor of $\sim$100 is very rare in AGN, but
has occasionally been seen. 
X-ray emission from the narrow-line Seyfert I galaxy (NLS1) WPVS007
dimmed by a factor 400 between 1990 and 1993 (\cite{Grupe95b}),
and has been faint ever since (\cite{Grupe08}).
Subsequent observations with FUSE revealed the development of a
strong BAL outflow (\cite{Leighly09}) which led Grupe et al. (2008) to
conclude that the strong X-ray variability and very soft spectrum
are caused by variations in a highly-ionized absorber.
Another NLS1, PHL~1092, dropped in flux by a factor 200, while
maintaining roughly the
same spectrum, interpreted as a reduction in the comptonising
medium (\cite{Miniutti09}). GSN~069, a low-luminosity Seyfert II,
increased in flux by a factor 200
over 10 years, giving a very soft, apparently thermal, spectrum, which
was interpreted as a transition in the disk state by 
Saxton et al. (2011).

All these objects are shown to be persistent AGN
by the strong narrow emission lines in their optical spectra.

The crucial arguments for why TDE candidates differ fundamentally from
the above high-variability AGN have been laid out before (notably in
\cite{Komossa99}, \cite{Komossa02} and \cite{Gezari09})
but are worth examining again in the light of recent
results.

\begin{itemize}
\item The absence of narrow lines in the optical spectrum. Limits on the
bolometric luminosity of any persistent AGN emission can be inferred
from the strength of the narrow emission lines, notably [OIII]
(e.g. \cite{Lamastra}), which gives the average $L_{bol}$ seen by the NLR
over the last $\sim10^{4}$ years and so is unaffected by temporary rises
or drops in the accretion rate or line-of-sight absorption.
On a note of caution, the emergence of
a subclass of AGN which are X-ray bright but show no optical features
(XBONGs; \cite{Comastri02})
means that very rare examples of AGN without optical emission lines do exist. 
Trump et al. (2009) discuss a sample of 48 XBONGs from the COSMOS
survey and show that thirteen have anomalously weak narrow
ionization lines given their bolometric flux. They suggest that
these AGN may be weakly accreting from a truncated accretion disk.
Another explanation for at least some XBONGs has been the dominance of a
host galaxy outshining faint emission features.  
While the physical nature of some XBONGs is still unclear, to our knowledge
no XBONG has yet been seen to display strong X-ray variability and so
the chance of a highly-variable AGN, which is also an XBONG, being mistaken
for a TDE is small.

Blazars also show rapid variability and weak emission lines and could
in principle be mistaken for TDE. They can usually be recognised by 
their characteristic radio emission and radio-to-optical spectral index
and have been successfully excluded in interpretations of previous
events (e.g. \cite{Gezari09}, \cite{Cenko}). 
\\

\item The X-ray flux should be dominated by the availability of accretable material
and should fall over the long-term as $t^{-5/3}$. Recent numerical
simulations show that while the initial decline depends on a number
of factors, the trend should tend to $t^{-5/3}$ within $\sim$1 year.
There is no reason why variable absorption or disk state transitions
should follow this law.
\\

\item The X-ray emission should continue to fall until it reaches levels
below those of even nearby classical AGN and should not rise again
on human timescales.{\footnote{Note, however, the possibility of repeat
tidal disruptions on the timescales of years or centuries, if the SMBH
at the galaxy core is a binary (\cite{Chen}), or if the SMBH is
off-centre due to gravitational wave recoil (\cite{KomossaMerritt08}).}}
\end{itemize}
It is the combination of these three tests which makes the TDE
interpretation
so strong for a small number of ROSAT detected candidates. NGC~5905
was optically consistent with an HII galaxy from ground-based
observations (\cite{Komossa99}) while with HST, a low-luminosity
[OIII] emission line (\cite{Gezari03}), conceivably induced by the flare, was detected.
The decline in flux of NGC~5905 fits well to a $t^{-5/3}$
curve (\cite{Komossa99}) and monitoring over 11 years shows L$_{X}$ falling
below $10^{39}$ \lumUnits to a level where the remaining emission is
probably
due to starburst activity (\cite{Halpern04}).

 
Finally, it should also be noted that the disruption of a star by a SMBH
with a pre-existing
accretion disk, may actually be more frequent than a disruption by
a naked SMBH (\cite{KarasSubr07}). However, it would be very difficult
to unambiguously identify such an event using the above conditions.
A case in point is that of IC~3599 (\cite{Brandt}; \cite{Grupe95a}).

\subsection{Could the flare of \sdss be an AGN event ?}


First, we note, that in none of our two optical spectra did we actually detect 
any emission lines which indicate the presence of a permanent AGN. 
From the WHT spectrum we infer $L_{bol}\leq3\times10^{41}$ ergs s$^{-1}$
for any persistent AGN activity and $L_{bol}=5-14\times10^{44}$ \lumUnits
during the \xmm slew observation.
For this flux change to be due to variable absorption a change in column
of $\Delta N_{H}>10^{23} cm^{-2}$ would be needed for a neutral absorber
and somewhat larger for an ionised absorber.
If variable absorption was the sole cause of the flux variations
seen in the X-ray light curve,
then a strong imprint of absorption would be seen in the 2010-12-23
\xmm observation. Evidence of strong absorption is not seen and so
we can exclude absorption effects as the primary cause of the flux
variations. 
And, again, if we only had line-of-sight absorption, the narrow emission
lines should still be present in the optical spectrum.  
A situation, where a thick neutral absorber has completely shielded
the whole NLR in all directions for the last millenia, but has now opened
a tiny hole, for a temporary, partially covered, glimpse into the central engine,
appears highly contrived. Neither does it explain
the peculiar X-ray spectrum that we observe (which is unlike any 
classical AGN), nor should it mimic the lightcurve decline law
expected from a TDE.   
Similar arguments have been given for the case of NGC\,5905 and
other TDEs before.

 From the relationship of black hole mass to bulge K-band luminosity
(\cite{MarconiHunt}) we find $3\times10^{5}<M_{BH}<2\times10^{7}M_{\odot}$
in the case of \sdssns,
where the range indicates the uncertainty
in the galaxy type and hence the contribution of star formation to the K
band flux.
The UV flux and template fits to the optical spectra imply that there is
little recent star formation in the galaxy, which in turn implies that 
the source resides in an early-type host galaxy, where the 
K-band luminosity will be dominated by disk emission. For this reason
we prefer the higher value of $M_{BH}=2\times10^{7}M_{\odot}$ 
with an error from the correlation of 0.3 dex.

 From the peak bolometric luminosity of $L_{bol}=5-14\times10^{44}$
\lumUnits
and using $L_{edd}=1.3\times10^{44}M_{6}$ \lumUnitsns,
(where $M_{6}$ is the black hole mass in units of $10^{6}$M$_{\odot}$),
we find an Eddington
ratio at the time of the \xmm slew observation of 0.2-33
and $<10^{-4}-10^{-2}$ for any persistent emission. This would then imply
that the AGN had undergone a sudden change from an inefficient accretion
state (e.g. advection dominated accretion flow [ADAF]; \cite{Narayan})
to an efficient
state (e.g. slimdisk; \cite{ShakSun73})
and had then returned to the inefficient state over $\sim1$ year.

\subsection{Could \sdss be a blazar or have jet emission?}

The strong limits on radio emission, discussed in section 5,
effectively preclude the presence of a strong jet in \sdssns. 
The radio-to-optical spectral index
$\alpha_{RO}=-log(f_{\nu,R}/f_{\nu,O} / log(\nu_{R}/\nu_{O})<0.14$,
where the radio is taken at 1.4 GHz and the optical from the
WHT spectrum at 4400\AA, is below the minimum value 
of $\alpha_{RO}=0.2$ associated with 
the blazar population (\cite{Beckmann03}).
In this regard \sdss is very different from 
\swtd where an on-axis jet appears to dominate the radio 
(\cite{Zauderer}; \cite{Berger}) and X-ray (\cite{Burrows11}; \cite{Bloom})
 emission.
If an {\it off-axis} jet is launched by the accretion of tidally disrupted 
material then it should radiate strongly in the radio band 
when it interacts with
the interstellar medium and a reverse shock is formed (\cite{Giannios}). 
The radio emission is predicted
to peak after $\sim 1$ year, at $\nu$=25 GHz, with a flux of 
$\sim2$ (D/GPc)$^{-2}$ mJy.
At the distance of \sdss (700 Mpc) the predicted fluxes 
at 1.4, 4.8 and 8.3 GHz are 0.4, 2.2 and 2.8 mJy respectively. 
The \sdss 3-sigma upper limits taken 
16 months after discovery are a factor 2, 16 and 26 below these values. 
This suggests that a jet was not launched during this event.

\subsection{Tidal disruption}

The gradual decline of the X-ray emission, albeit with large excursions, and
the relatively constant spectral shape are arguments in favour of a
supply of material which is being exhausted at the predicted rate
of $t^{-5/3}$ (\cite{Phinney}). 
Lodato, King \& Pringle (2009) predicted that the
decline in flux would depend on the
internal state of the star, and would usually be more gradual initially,
tending to $t^{-5/3}$ only after several months. With the strong
 variations
seen in this lightcurve it is impossible to distinguish between the two
models.\footnote{Together with SWIFT~J164449.3+573451 and SWIFT~J2058.4+0516
(\cite{Cenko11b}) this
is the best sampled X-ray light curve of a TDE candidate made to date.
Earlier
works have not made the critical observations in the weeks and months
following the flare detection.}
 
The light curve resembles that of \swtdns, suffering continual
strong variability on a timescale of days. In \swtd this has been
ascribed to the precession of a relativistic jet which has been 
switched on by a tidal disruption event (\cite{Burrows11}).
Here we need to search for a different mechanism involving flaring 
and/or absorption processes. The rate of return of tidal debris is
likely to be irregular and an accretion structure
formed from this material may well exhibit X-ray variability in excess of that
seen from established AGN accretion disks.

The X-ray spectrum of \sdss is not the thermal radiation expected in the
early stages of a disruption, if the density is sufficient to thermalise
the photons (\cite{Rees88}) and neither
is it the typical signature of an accretion disk, seen in the late stages
of several other TDE candidates (\cite{Komossa04}; \cite{Vaughan},
\cite{Esquej08}).
It is best fit with a \brem or broken power-law model which becomes
softer with time and/or reducing flux.

 
\paragraph{Where is the thermal component?}

The expectation was that tidal disruption events would, at least initially,
have a purely thermal spectrum, either from a thin disk (\cite{Cannizzo})
or a thick disk (\cite{Ulmer99}). The effective temperature ($T_{eff}$)
of the emission
will be mainly dependent upon the inner radius assumed in the disk
configuration. For Eddington-limited, thick-disk accretion, \cite{Ulmer99}
showed that $T_{eff}\approx40M_{6}^{-1/4}$ eV.
In the thin-disk case $T_{eff}\approx56M_{6}^{-1/4}$ eV while
if the debris forms at the tidal disruption radius then
$T_{eff}\approx24M_{6}^{-1/12}$ eV (\cite{Komossa02}).
Thermal emission, from the inner edge of the accretion disk,
should also exist in persistent AGN. 
Very rarely it can be seen if the AGN
has a small BH mass and lacks a comptonisation region; 
e.g., RX~J1633+4718 (\cite{Yuan}) and GSN~069 (\cite{Saxton11}).

\begin{figure}
\centering
\rotatebox{0}{\includegraphics[height=6.5cm]{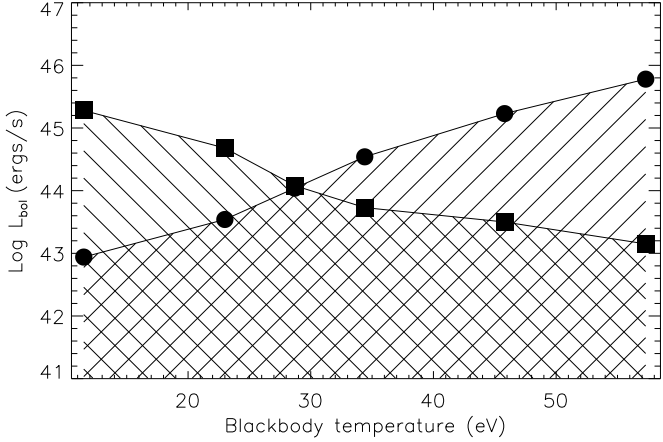}}
\caption[Black body limits]
{ \label{fig:bbodylimit} Limits on the bolometric luminosity of
blackbody emission from \sdss during the observation of 2010-06-22.
Filled squares represent limits from soft X-ray fits while filed circles
give
the limits from the UV emission. The cross-hatched area shows the allowable
L$_{bol}$ as a function of the rest frame blackbody temperature.}

\end{figure}

Although most of the previous X-ray selected TDE candidates have shown
soft X-ray spectra, none of the observations, at the peak of the disruption,
 have had sufficient spectral resolution and statistics to conclusively
constrain the spectral model.
In the \xmm observation of 2010-06-22, which appears to be close in time
to the peak of the flare, we ought to see black-body emission with
a temperature of a few tens of eV and $L\sim10^{44}M_{6}$ \lumUnits
(\cite{Rees88}).
Such emission is not obviously present in the spectrum that we observe.
A lower limit on the temperature of a putative black-body component
can be found from the UV flux. From the upper limit to the difference
in the U filter flux between the 2010-06-22 and 2010-12-23
\xmm pointed observations
(0.025 count s$^{-1}$), a $L_{bol}=10^{44}$ \lumUnits black-body
component must have kT$>29$ eV.
 From X-ray spectral fits we can put a limit
of F$_{0.2-2}<2.4\times10^{-13}$\fluxUnits
on an additional soft component in the 2010-06-22 \xmm pointed observation.
In Fig.~\ref{fig:bbodylimit} we show that the UV and X-ray results
together constrain
the bolometric luminosity of any optically thick thermal emission
to $L_{bb}\leq10^{44}$
 \lumUnitsns, even if the peak of the emission lies in the EUV
band, during the 2010-06-22 observation.

\paragraph{Absorption mechanisms:}

The strong decrease in flux between 20 and 40 days after discovery 
(Fig.~\ref{fig:lcurve}) may be due to absorption by line of sight material.
As a guide, the observed flux drop between the \swift 
observations of 2010-06-30 and 2010-07-07 requires a neutral absorber
with $N_{H}>5\times10^{22}$ cm$^{-2}$.
A possible source of this absorption could be material driven 
from the system by radiation
pressure during an early super-Eddington accretion phase (\cite{Strubbe09}).
This is calculated to produce soft X-ray absorption with an attenuation 
which is dependent on M$_{BH}$ and the expansion
velocity (\cite{Strubbe11}) and timescales broadly compatible
with those that we see in \sdssns, i.e. apparently strong absorption 
starting after 20-30 days which has greatly reduced after 130 days.
The expanding envelope would itself, however, emit thermally with 
$L_{bol}\sim10^{44}$ M$_{6}^{11/9} (t/day)^{-5/9}$ \lumUnits 
and a temperature of $\sim 10-20$ eV. This is not really
compatible with the limits on thermal emission discussed in 
the previous section. We note that the optical emission from the
envelope, $L_{opt}\sim10^{41-43}$ \lumUnits would not 
necessarily be detectable in 
this case against the background emission from the galaxy.

Another possible source of absorbing material is the
non-bound stellar debris (\cite{Khokhlov}), even though a special
viewing angle is required in this case.

\paragraph{What are we seeing?}

While we clearly aren't observing optically thick thermal radiation,
the 2010-06-22 \xmm pointed spectrum is well fit with optically thin
emission at a temperature of kT$\sim390$ eV. Emission at this temperature
should produce strong emission lines from OVII, OVIII, NVI and NVII.
No lines are seen and a fit with an {\em APEC} model gives an upper
limit to the metal abundance of 0.5\% of solar values.
This effectively rules out the possibility that we are observing emission
from debris heated as it falls back and before it circularises.
Khokhlov \& Melia (1996) predict that unbound disrupted material will be expelled
from the system in a fan-like shape at high velocity. X-rays are created
when this collimated beam shocks against circumnuclear material.
The resulting spectrum should be much like that of an extremely energetic
supernova remnant, typically modelled as a vpshock model. This
model doesn't fit well, as it also overpredicts line strengths.
Also it is not clear why the
emission from the shock would effectively turn off after 20 days.

If the returning material quickly settles into an accretion disk
(\cite{Cannizzo}; \cite{Ulmer99}) then the X-ray spectrum could be due
to some form
of comptonisation of thermal photons by hot electrons. Nevertheless,
this would have to occur without the optical and/or UV
flux increasing significantly.

\begin{figure}
\centering
\rotatebox{270}{\includegraphics[height=9cm]{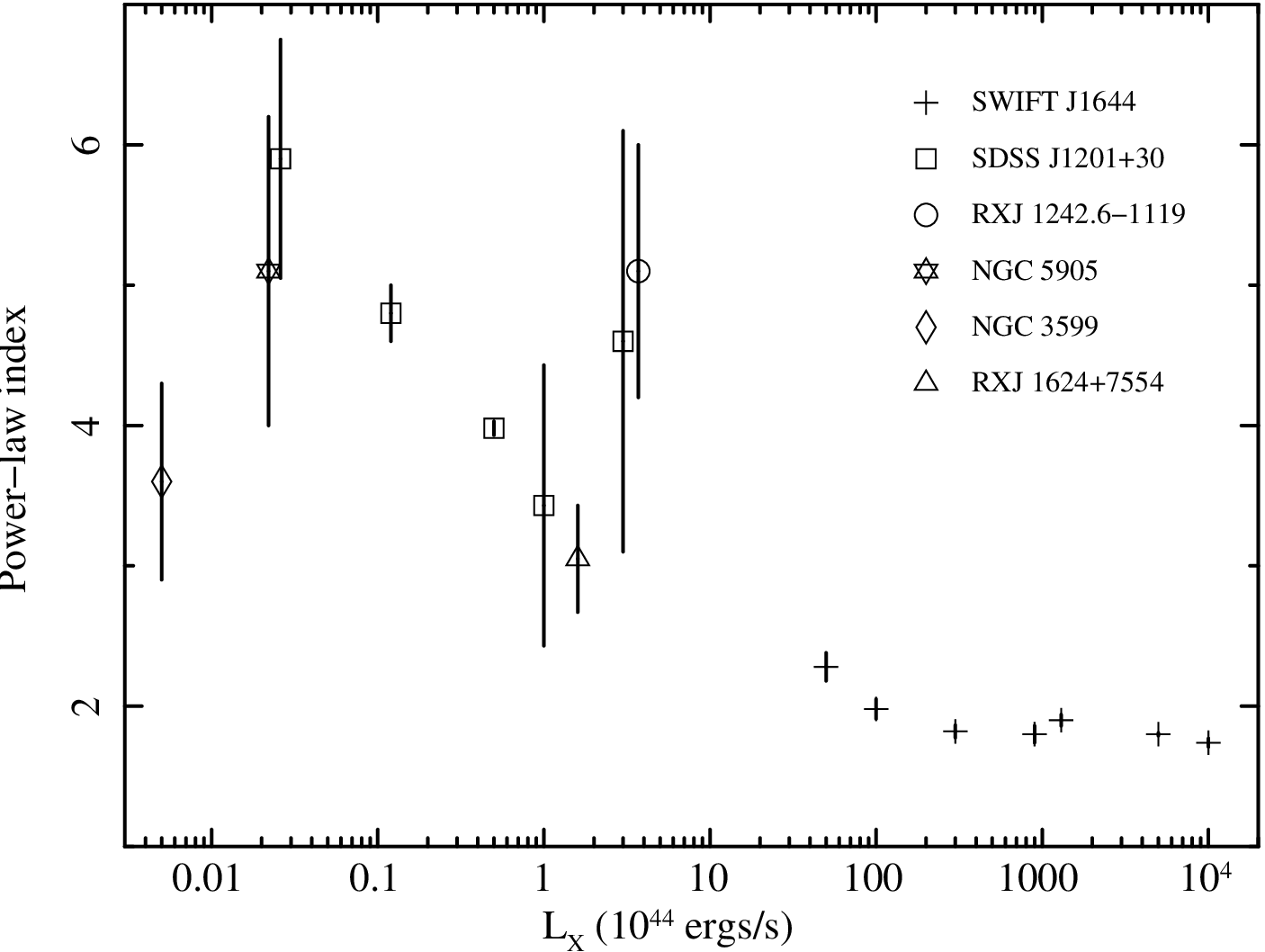}}
\caption[Relationship of spectral slope with X-ray luminosity]
{ \label{fig:specflux} The spectral slope of a single power-law,
absorbed by the galactic column and intrinsic absorption,
plotted against X-ray luminosity:
\swtd (\cite{Burrows11}), \sdss and NGC~3599 (this work),
NGC~5905 (\cite{Bade96}),
RXJ~1242.6-1119 (\cite{Komossa99b}; this fit only uses Galactic
absorption),
RXJ~1624+7554 (\cite{Grupe99})}.
\end{figure}

While the absence of radio emission in \sdss makes it unlikely that 
it shares the same emission mechanism as \swtdns, the similarities
in the light curves are intriguing. In fact the relationship of the
spectral hardness with X-ray flux (Fig.~\ref{fig:hratio}) goes in the 
same direction in the two sources.
In Fig.~\ref{fig:specflux}
we compare the evolution of the X-ray spectral hardness (as 
parameterised by the power-law slope) with X-ray luminosity
for \sdss and \swtd and compare it with a selection of earlier TDE 
candidates in the early peak phase of their emission.



\subsection{Reprocessed optical radiation}
The two optical observations allow constraints to be placed on the amount
of reprocessing material in the vicinity of the SMBH.
The intense radiation from the peak of the flare will naturally be
reprocessed by the stellar debris; predictions give an $H{\alpha}$
line, which is broad and variable on timescales of hours to days, with a
luminosity of $<10^{37-39}$ \lumUnits (\cite{Bogdanovic}).
The Calar-Alto/TWIN spectrum, taken 12
days after discovery, has an $H{\alpha}$ limit
$L_{H{\alpha}}<10^{39}$ \lumUnitsns,
and so we are not sensitive 
to these broad lines
for sources with spectra like \sdss  (note its
Hydrogen absorption features in the optical spectrum)
for 1 hour exposures on a 4m class telescope.

If the source were an AGN with a persistent luminosity
of L$_{bol}<3\times10^{41}$ \lumUnits then the radius of the BLR
would be 1-2 light days (\cite{Bentz})
and the optical spectrum of 2010-06-22 should show strong broad lines
from this region.
In practise an AGN appears to need a minimum strength
L$_{bol}>10^{42}$ \lumUnits to produce a BLR (\cite{Nicastro}) and so
we would perhaps not expect to see broad optical lines in this case.

The WHT observation in May 2011 probed a light cone which can be
approximated by a wave front 1-2 light months wide (when the source was
near its peak)
lasting from June to July 2010. We will see reprocessing in the WHT
observation from material at 5-11 light months distance from the nucleus.
For $M_{BH}\sim10^{7}M_{\odot}$, 5-11 light months corresponds to the
coronal line region. A close analogy to this situation was
found in SDSS~J0952+2143 (\cite{Komossa09}) where very strong
(L$\sim10^{40}$ \lumUnitsns)
emission was seen from FeVII, FeX, FeXI and FeXIV. In that case the
unseen X-ray flare was calculated to have parameters similar to the ones
seen in \sdssns; $L_{X}\sim10^{44}$ \lumUnits and
$M_{BH}\sim10^{7}M_{\odot}$.
A key factor in the production of coronal lines
is the soft X-ray flux although the line strength will also 
depend on the UV continuum. While detailed modelling is beyond the
scope of this paper, we note that the typical 
CLR line luminosity is $\approx0.1\%$ of the observed X-ray luminosity
in AGN (\cite{Nagao}). We might therefore expect to see Fe line
strengths of L$_{Fe}\sim10^{40-41}$ \lumUnitsns. The limits on emission
from these narrow lines in the WHT observation are  L$_{Fe}<10^{39}$
\lumUnits
suggesting that the total mass of flourescing material in the CLR
of \sdss is $<10\%$ of that of a typical AGN or that of SDSS~J0952+2143.

\section{Summary}

A soft X-ray flare was seen in an \xmm slew observation
of the galaxy \sdss in June 2010
with an implied $L_{bol}=5-14\times10^{44}$ \lumUnitsns. The galaxy shows
no evidence for previous AGN activity. Together with the absence of strong
intrinsic absorption in subsequent X-ray observations, this makes it
unlikely
that the flare is due to a change in line-of-sight obscuring material
towards a persistent AGN. Large variability, on timescales of a week,
is apparent in the light curve but the general downward trend, over a year,
is not inconsistent with the reduction in accretable material expected
from the disruption of a stellar object by a SMBH. 

The strong variability seen in the light curve is likely due to a 
combination of flaring and absorption events, perhaps related to 
clumpy accretion.  
The X-ray spectrum is not that of optically thick
thermal emission or that of a typical accretion disk. It can be modelled
by \brem or a double power-law, which becomes softer as the flux
decreases. 

Optical spectra indicate an absence of material in the BLR and
CLR of the nucleus of this galaxy. Deep radio upper limits imply
that a jet was not launched during this event.

This first result shows that a monitoring program with \swift and \xmm
of flaring extragalactic sources is capable of
producing X-ray light curves and spectra of high quality, 
which in this case are 
not simply explained by current models.


\acknowledgements
Funding for the SDSS and SDSS-II has been provided by the Alfred P. Sloan Foundation, the Participating Institutions, the National Science Foundation, the U.S. Department of Energy, the National Aeronautics and Space Administration, the Japanese Monbukagakusho, the Max Planck Society, and the Higher Education Funding Council for England. The SDSS Web Site is http://www.sdss.org/.
The National Radio Astronomy Observatory is a facility of the National Science Foundation operated under cooperative agreement by Associated Universities, Inc.
We thank Calar Alto Observatory and NRAO for allocation of director's 
discretionary time to this programme. This research has been funded 
by Spanish grants AYA 2010-21161-C02-02, CDS2006-00070 and 
PRICIT-S2009/ESP-1496.
We thank the XMM OTAC for approving this program.
The XMM-Newton project is an ESA science mission with instruments and contributions directly funded by ESA member states and the USA (NASA).
The \xmm project is supported by the Bundesministerium f\"{u}r Wirtschaft 
und Technologie/Deutches Zentrum f\"{u}r Luft- und Raumfahrt i
(BMWI/DLR, FKZ 50 OX 0001), the Max-Planck Society and the Heidenhain-Stiftung.
We thank the \swift team for approving and performing the monitoring 
observations. This work made use of data supplied by the UK \swift Science Data Centre at the University of Leicester.
AMR acknowledges the support of STFC/UKSA/ESA funding.

\end{document}